\documentclass{aa}
\usepackage[varg]{txfonts}
\usepackage{graphicx}
\usepackage{txfonts}
\usepackage{lipsum}
\usepackage{subcaption}         % necessary for continued figures, example in section 3
\usepackage{lscape}             % to rotate a single page table, example in appendix.
\usepackage{placeins}           % useful with \FloatBarrier, to keep 
\usepackage[colorlinks=True, linkcolor=green, citecolor=blue, urlcolor=red]{hyperref}

\begin{document}

\title{X-ray Quasi-Periodic Oscillations in Active Galactic Nuclei and Their Implications for the Changing Look Phenomenon}

\author{Mouyuan Sun\inst{1}
  \and Shuying Zhou\inst{1}
  \and Jihong Liu\inst{1}
  \and Ning Jiang\inst{2,3}
  \and Zhen-Yi Cai\inst{2,3}
  \and Hai-Cheng Feng\inst{4,5,6,7}
  \and Hengxiao Guo\inst{8}
  \and Zhi-Xiang Zhang\inst{9}
  \and Qinbo Han\inst{1}
  \and Juan Li\inst{1}
  \and Linyue Jiang\inst{1}
  \and Yu-Jing Xu\inst{1}
  \and Junfeng Wang\inst{1}
  \and Jun-Xian Wang\inst{2,3}
  \and Yongquan Xue\inst{2,3}}

\offprints{M. Sun, \email{msun88@xmu.edu.cn}}

\institute{Department of Astronomy, Xiamen University, Xiamen, Fujian 361005, People’s Republic of China 
 \and Department of Astronomy, University of Science and Technology of China, Hefei, Anhui 230026, People’s Republic of China 
 \and School of Astronomy and Space Science, University of Science and Technology of China, Hefei, Anhui 230026, People’s Republic of China 
 \and Yunnan Observatories, Chinese Academy of Sciences, Kunming 650216, Yunnan, People's Republic of China 
 \and Key Laboratory for the Structure and Evolution of Celestial Objects, Chinese Academy of Sciences, Kunming 650216, Yunnan, People's Republic of China 
 \and Center for Astronomical Mega-Science, Chinese Academy of Sciences, 20A Datun Road, Chaoyang District, Beijing 100012, People's Republic of China 
 \and Key Laboratory of Radio Astronomy and Technology, Chinese Academy of Sciences, 20A Datun Road, Chaoyang District, Beijing 100101, People's Republic of China 
 \and Shanghai Astronomical Observatory, Chinese Academy of Sciences, 80 Nandan Road, Shanghai 200030, People’s Republic of China 
 \and College of Physics and Information Engineering, Quanzhou Normal University, Quanzhou, Fujian 362000, People’s Republic of China}

\date{Received xxx}

\abstract{X-ray timing of active galactic nuclei (AGN) provides a unique probe of gas accretion onto supermassive black holes (SMBHs). Quasi-periodic oscillations (QPOs), which trace gas dynamics in the strongly curved spacetime around SMBHs, are rare in AGN. These signals often are analogs of high-frequency QPOs occasionally seen in some black-hole X-ray binaries, and their scarcity in AGN can partly be attributed to the low frequencies expected for typical SMBH masses. Intriguingly, robust X-ray QPO detections in SMBH systems have so far been reported only in narrow-line Seyfert 1 galaxies (NLS1s) and tidal disruption events (TDEs). Here we report the discovery of a QPO candidate during the 2018 outburst of the changing-look AGN (CL-AGN) NGC 1566. Numerical simulations indicate that the disk epicyclic oscillations responsible for high-frequency QPOs are damped by magnetohydrodynamic turbulence unless the accretion flow is misaligned and/or eccentric. In TDEs, the stellar debris stream is naturally misaligned with the SMBH spin, while NLS1s may host misaligned disks due to their youth. Motivated by the QPO candidate in NGC 1566, we propose that CL-AGN accretion is also misaligned—potentially fueled by captured, free-falling broad-line region clouds. This model naturally explains why CL-AGN transition timescales are much shorter than the standard disk viscous timescale. This picture can be tested by searching for QPOs or quasi-periodic eruptions in other CL-AGN.} 


\titlerunning{X-ray QPOs and CL-AGN}
\authorrunning{M. Sun et al.}
\maketitle
\nolinenumbers

\section{Introduction} \label{sec:intro}
The timing properties of active galactic nuclei (AGN) provide unique insights into the structure and evolution of their central engine---the supermassive black hole (SMBH) and its accretion disk---which are otherwise too compact to be spatially resolved. For instance, stochastic variations in X-ray and UV/optical continua---likely driven by MHD turbulence \citep[e.g.,][]{Balbus1998, Noble2009, Sun2020-char} in the corona or accretion disk---are often highly correlated. The variable ionizing continuum can photoionize high-velocity clouds around SMBHs, producing broad emission lines with time-varying fluxes, or be absorbed by a dusty torus and re-emitted in the mid-infrared (MIR) bands. Reverberation mapping \citep{Blandford1982} measures time lags between these flux variations, thereby constraining the emission-region sizes \citep[for a recent review, see, e.g.,][]{Cackett2021} and enabling the estimation of SMBH masses ($M_{\mathrm{BH}}$) across cosmic time. 

Timing studies reveal significant challenges to our understanding of AGN. On short timescales (hours to days), an interesting problem is the rarity of X-ray quasi-periodic oscillation (QPO) detections in AGN \citep[e.g.,][]{Vaughan2012}. QPOs with different frequencies are observed in various spectral states of black-hole X-ray binaries (BHXRBs), serving as powerful probes of accretion physics around stellar black holes \cite[e.g.,][]{Remillard2006, Ingram2019}. If the accretion physics is scale-invariant from stellar black holes to SMBHs, AGN should exhibit QPOs beyond stochastic variability in long X-ray exposures. For instance, counterparts to high-frequency BHXRB QPOs---probing the gas dynamics near the innermost stable orbits---should be detectable with tens of kiloseconds of \textit{XMM-Newton} observations. The identification of AGN analogs of low-frequency QPOs ($0.1$--$30$ Hz) requires very long X-ray light curves. Over the past two decades, there have been only a few robust AGN QPO detections \citep[e.g.,][]{Gierlinski2008, Jin2021, Xia2024, Zhang2025}, possibly due to observational limitations and red noise contamination \citep[e.g.,][]{Vaughan2005, Vaughan2013}. \cite{Vaughan2012} systematically analyzed the \textit{XMM-Newton} light curves of $104$ AGN and only identified one source \citep[RE J1034+396, which is firstly discovered by][]{Gierlinski2008} with QPO signals. Nevertheless, QPOs are detected \citep[e.g.,][]{Pasham2019, Masterson2025, ZhangWJ2025} in five or more tidal disruption events \citep[TDEs; see, e.g.,][]{Rees1988}, which are extra-galactic transients that occur when a star is tidally disrupted and subsequently accreted by an SMBH. Hence, the QPO detection rate seems to be unusually high in TDEs because only several dozen TDEs have been monitored with intensive X-ray observations \citep[e.g.,][]{Gezari2021}. 

On long timescales (months to years), AGN show unexpected ``changing-look'' behavior \citep[e.g.,][]{Alloin1986, MacLeod2016, Guo2019, Guo2024, Dong2025}, where their broad emission lines disappear and/or reappear. Multi-wavelength variability studies confirm that these changing-look AGN (CL-AGN) undergo large accretion-rate variations---not dust-extinction changes---on timescales of years \citep[for a recent review, see][]{Ricci2023}. The variability timescales of CL-AGN are two to three orders of magnitude shorter than the predicted viscous timescales \citep[e.g.,][]{Lawrence2018}. This inconsistency presents a serious challenge to the standard accretion disk model \citep{SSD}, which successfully explains the thermal spectral state of BHXRBs \citep{Remillard2006}. This discrepancy may indicate strong magnetic field effects \citep[e.g.,][]{Dexter2019,Feng2021, Ma2025}, disk instabilities \citep[e.g.,][]{Sniegowska2020}, or a distinct gas accretion process in AGN \citep[e.g.,][]{Liu2021}. Together, these two problems highlight our poor understanding of time-dependent SMBH accretion \citep[e.g.,][]{Antonucci2013} on short and long timescales. 

The two problems could be well related. High-frequency QPOs in BHXRBs are found exclusively in the steep power-law (SPL) state with a strong steep power-law spectral component. The SPL state typically occurs around the peak of a BHXRB outburst \citep[e.g.,][]{Remillard2006}, is associated with a large Eddington ratio and transient jet ejections \citep[e.g.,][]{Massi2011}, and is presumably powered by an unsteady accretion-flow structure. Thus, could the analogs of high-frequency QPOs in AGN also be associated with unusual accretion processes with unstable gas inflow? A number of models have been proposed to explain low-frequency and high-frequency QPOs in XRBs \citep[e.g.,][]{Stella1999, Abramowicz2001, LiLX2004, Kato2008, You2018}. Interestingly, recent Magnetohydrodynamic (MHD) simulations \citep{Dewberry2020} suggest that MHD turbulence driven by magnetorotational instability, which is critical for removing gas angular momentum \citep[e.g.,][]{Balbus1998}, can effectively damp the inertial oscillations \citep[i.e., the $g$-mode; see, e.g.,][]{Kato2008} in the very attractive high-frequency QPO model, the discoseismic model \citep[for a brief discussion, see, e.g.,][]{Smith2021}. If this is the case, then other mechanisms involving warped and/or eccentric gas inflows would be required to excite \citep[e.g.,][]{Kato2004, Ferreira2008, Henisey2009} and maintain inertial oscillations \citep[see, e.g.,][]{Dewberry2020, Musoke2023, Bollimpalli2024} at the radial epicyclic frequency. Might such mechanisms also provide clues to our understanding of long-term AGN variability (e.g., the CL phenomenon)? 

Here we are motivated to search for possible X-ray high-frequency QPOs in the repeating CL-AGN NGC 1566 \citep[e.g.,][]{Alloin1986, Oknyansky2019, Ochmann2024}. This source is an ideal target for the following reasons. First, the CL transitions are likely driven by large accretion-rate fluctuations, which may also amplify inertial oscillations in the innermost regions. Second, a millihertz QPO is reported in the outburst of the CL-AGN 1ES 1927+654. Third, NGC 1556 is frequently observed by \textit{XMM-Newton} and other X-ray facilities \citep[e.g.,][]{Jana2021} during its 2018 outburst \citep[e.g.,][]{Oknyansky2019}. Fourth, NGC 1566 shows significant Fe~\textsc{II} emission during the 2018 outburst, indicating a large peak Eddington ratio. Estimates of this peak Eddington ratio range from $6\%$ to $ 12\%$, depending upon the SMBH mass \citep[e.g.,][]{Jana2021, Ochmann2024}. Furthermore, this source is classified as a narrow-line Seyfert 1 (NLS1) by \cite{Xu2024}, a class often considered the AGN analogy of BHXRBs in the SPL state. Fifth, given its SMBH mass $M_{\mathrm{BH}}=4\sim 8\times 10^6\ M_{\odot}$ \citep{Ochmann2024}, the expected period of the high-frequency QPO, which should be $\simeq 2000\sim 4000$ s according to the scaling relation of BHXRBs \citep{Remillard2006}, can be probed by \textit{XMM-Newton} observations. The manuscript is formatted as follows: Section~\ref{sec:method} describes the QPO detection methodology; Section~\ref{sec:results} presents the QPO detection result; Section~\ref{sec:dis} discusses the physical implications. The main conclusions are summarized in Section~\ref{sec:sum}.

\section{Data and QPO Identification} \label{sec:method}
\subsection{Light Curve Extraction}\label{sec:data}
We are motivated to search for QPOs in the 2018 outburst of the repeating CL-AGN NGC 1566 \citep[e.g.,][]{Alloin1986, Oknyansky2019, Jana2021, Ochmann2024}. From November 2015 to August 2019, NGC~1566 was observed five times by \textit{XMM-Newton}. Here we focus on \textit{XMM-Newton} observations \citep[for a summary, see, e.g., table 1 of][]{Jana2021} with \textit{obs-IDs} of $0800840201$ (2018-06-26), $0820530401$ (2018-10-04), and $0840800401$ (2019-06-05) which are the three observations with the highest count rates and longest exposure time among the five exposures. The exposure times for the observations $0800840201$, $0820530401$, and $0840800401$ are $94$ ks, $108$ ks, and $94$ ks, respectively. We do not consider \textit{NuSTAR} observations because their count rates are too low to probe QPOs. We locally run \texttt{SAS} (version:22.1.0) with the latest calibration files to reduce the data and extract the $0.2$--$10$\,keV light curves. We process the raw observation data files using the \texttt{epproc} task to obtain calibrated EPIC event files. Next, we apply the \texttt{evselect} and \texttt{tabgtigen} tasks to select good time intervals without background flares (\texttt{RATE}$<=2$ within $10\ \mathrm{s}$ in the $10$-$12\ \mathrm{keV}$ band). Additionally, only single- and double-pixel events are retained (\texttt{PATTERN}$<=4$ and \texttt{FLAG}$==0$). The source area selects a $30''$ radius circle around the center of the source. The background area selects a $50''$ radius circle that is free of sources. We extract the source and background light curves using the \texttt{evselect} task; then, the \texttt{epiclccorr} task generates the background-removed light curves. The generated light curves for \textit{obs-IDs} of $0800840201$, $0820530401$, and $0840800401$ are rebinned in time bins of $100\ \mathrm{s}$. This time resolution is sufficiently small for our interested periods. 

\subsection{QPO Identification}\label{sec:psd}
Robustly searching for QPOs in AGN is challenging. AGN variability resembles ``red'' noise \citep[for a review, see e.g.,][and references therein]{Vaughan2013}, which can mimic periodic signals \citep[e.g.,][]{Vaughan2005}. Several methods have been proposed for QPO detection, such as the standard Fast Fourier Transform (FFT), and the popular Lomb-Scargle periodogram \citep[e.g.,][]{Scargle1982}, which can deal with irregular and gappy light curves, and the weighted wavelet Z-transform \citep[e.g.,][]{Foster1996}. Due to the stochastic nature of AGN variability, computationally heavy Monte Carlo resampling is often adopted to estimate the statistical significance of potential QPO signals. 

Recently, Gaussian Processes (GPs) have been introduced to model AGN light curves and search for QPOs in the time domain \cite[e.g.,][]{celerite, Hubner2022a, Hubner2022b, Yu2022-DHO, Zhang2023}. The GP method is particularly powerful in dealing with irregularly sampled light curves with heterogeneous measurement uncertainties. This approach assumes AGN flux variations are a stochastic process and follow a multivariate Gaussian distribution. By optimizing the covariance (kernel) function of the light curve, GPs model the underlying AGN variability and obtain the QPO parameters \cite[for more details, see, e.g.,][and references therein]{Hubner2022b}. Moreover, the GP method can also self-consistently model non-stationary signals, including but not limited to outburst light curves, by introducing the long-term mean functions \citep[e.g.,][]{Hubner2022a, Hubner2022b}. This advantage is particularly relevant for probing X-ray QPOs in the 2018 outburst of NGC 1566, which show a fast rise but slow decay \citep[figure 1 in][]{Jana2021}. 

Following \cite{Hubner2022a}, we model the potential QPO signal (hereafter the DRW+QPO model) using the following composite GP kernel function:
\begin{equation}
    \label{eq:full}
    k_\mathrm{QPO+DRW}(\Delta t) = k_\mathrm{DRW}(\Delta t) + k_\mathrm{QPO}(\Delta t),
\end{equation}
with 
\begin{equation}
    \label{eq:qpo}
    k_\mathrm{QPO}(\Delta t) = A \exp\left(-\frac{\Delta t}{\tau_{\mathrm{QPO}}}\right) \cos\left(\frac{2\pi \Delta t}{P_{\mathrm{QPO}}}\right),
\end{equation}
where $A$ is the QPO amplitude, $P_{\mathrm{QPO}}$ is the QPO period, $\tau_\mathrm{QPO}$ is the decay timescale, and $k_\mathrm{DRW}(\Delta t)$ is the kernel function of a damped random walk (DRW) process: 
\begin{equation}
    \label{eq:drw}
    k_\mathrm{DRW}(\Delta t) = \sigma^2 \exp\left(-\frac{\Delta t}{\tau_{\mathrm{DRW}}}\right),
\end{equation}
where $\Delta t$ is the time interval between two observations, $\sigma$ is the variability amplitude, and $\tau_\mathrm{DRW}$ is the damping timescale. This DRW+QPO model is fitted to the merged and rebinned \textit{XMM-Newton} light curve. 

While AGN stochastic variability is often described by a DRW process \citep[e.g.,][]{Kelly2009}, there is evidence indicating more complex GP kernel functions \citep[e.g.,][]{Kasliwal2015}. Here, we set the null hypothesis model (hereafter the noise model) by fixing $P_{\mathrm{QPO}}$ in Eq.~\ref{eq:full} to be an extremely large value (i.e., $10^6$ s). That is, this special kernel is identical to a summation of two DRW processes with different parameters and can fit the non-DRW noise. Just like the DRW+QPO model, the noise model is also fitted to the same \textit{XMM-Newton} light curve. 

\begin{figure*}
    \centering
    \includegraphics[width=17cm]{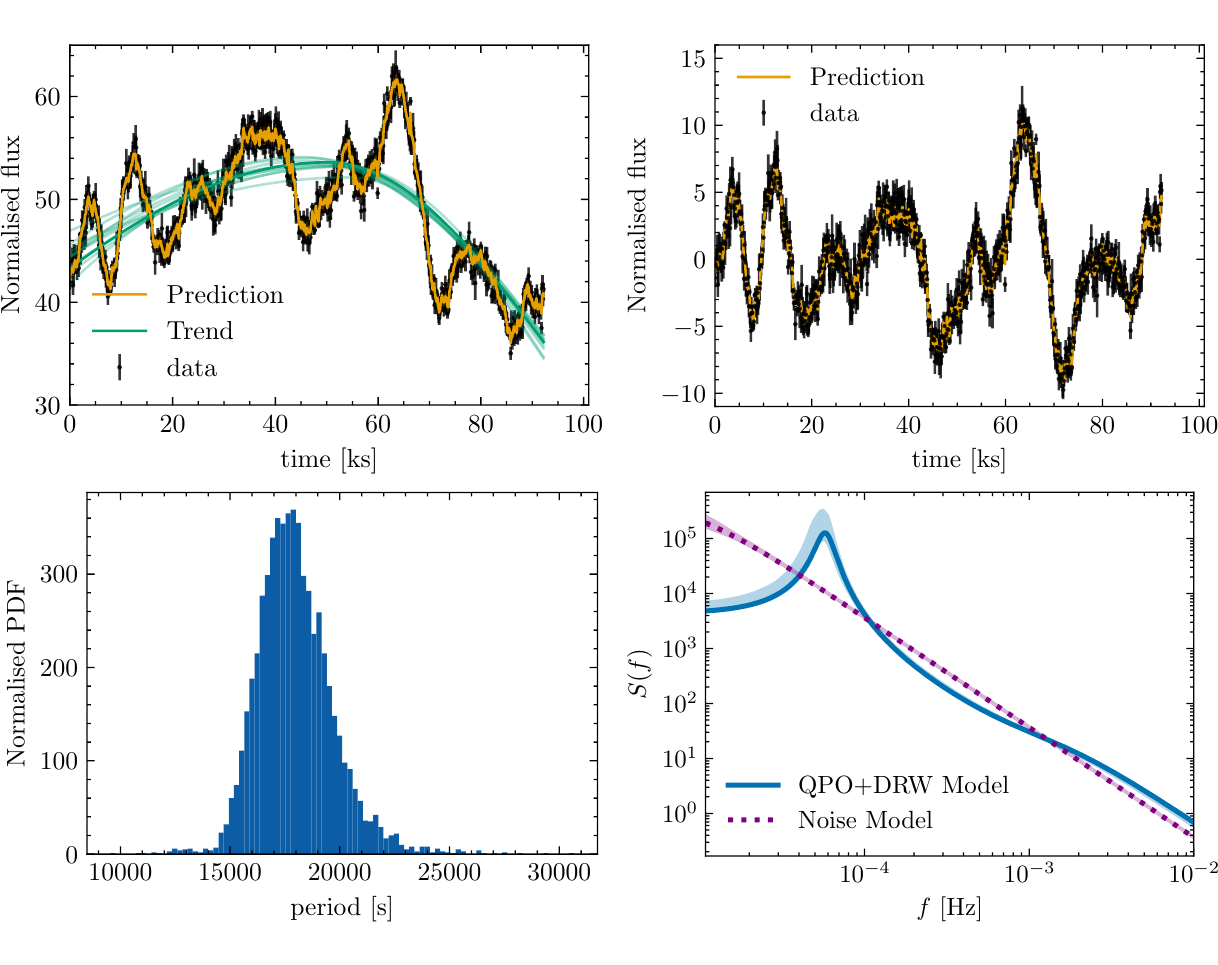}
    \caption{The GP fitting results of the QPO+DRW model for NGC~1566. Top-left: the background-subtracted \textit{XMM-Newton} EPIC light curve of NGC~1566 in the $0.2$--$10$\,keV band (black dots with error bars), binned at $100$ s. The light curve corresponds to the obs-ID $0800840201$, for which NGC 1566 was in the outburst phase. The green curves represent the long-term trend modeled by an shewed Gaussian function, which well describes the outburst profile. The yellow curve is the sum of the best-fitting QPO+DRW model and the trend. Top-right: the same as the top panel, but with the long-term trend (i.e., the green curve in the top panel) being subtracted. Bottom-left: the posterior distribution of the QPO period, yielding $P_{\mathrm{QPO}}=(1.78^{+0.17}_{-0.14})\times 10^4\ \mathrm{s}$. Bottom-right: the best-fitting PSDs for the QPO+DRW model (blue curve) and the noise model (purple dots); the shaded regions indicate the $1\sigma$ confidence intervals.} 
    \label{fig:fit}
\end{figure*}

As pointed out by \cite{Hubner2022b}, the ``mean'' function, which describes the long-term and/or non-stationary trend in a light curve, plays an important role in identifying QPOs. Again, following \cite[][see their section 2.3]{Hubner2022b}, we consider the five long-term mean functions, i.e., constant, polynomial functions, skewed Gaussian functions, skewed exponential functions, and fast-rise exponential decay (FRED) functions. Then, we use the \texttt{QPOEstimation}\footnote{For the code implementation, see \url{https://github.com/MoritzThomasHuebner/QPOEstimation}} package \citep{Hubner2022b} to fit the rebinned \textit{XMM-Newton} light curve of NGC 1566, obtaining the corresponding Bayesian evidence \citep[see Eq.~11 in][]{Hubner2022a} and best-fitting kernel parameters for the noise model and the DRW+QPO model. For each model, all five mean functions are considered. 

We compare the DRW+QPO and noise models using the Bayes factor \citep{Zhu2020, Hubner2022a}. The natural logarithmic Bayes factor is \citep{Hubner2022b},
\begin{equation}
    \ln \mathcal{BF} = \ln(\mathcal{Z}_\mathrm{QPO+DRW})-\ln(\mathcal{Z}_\mathrm{noise}),
\end{equation}
where $\mathcal{Z}_\mathrm{QPO+DRW}$ and $\mathcal{Z}_\mathrm{noise}$ are the Bayesian evidence of the DRW+QPO and noise models, respectively. The Bayes factor is a measure of the relative evidence provided by the light curve for the QPO+DRW over the noise models. Following previous works \citep{Kass1995, Zhu2020}, we treat $\ln\mathcal{BF}>3$ as strong and $\ln\mathcal{BF}>5$ as very strong support for the QPO+DRW model over the noise model. 

\begin{figure}
    \centering
    \includegraphics[width=8cm]{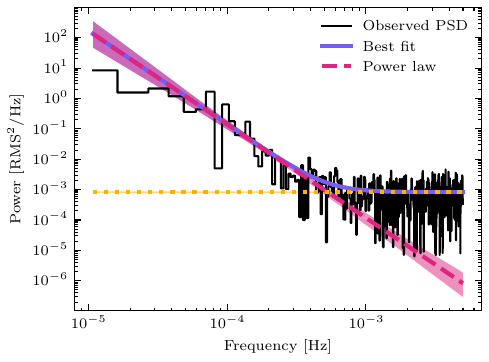}
    \caption{The observed and best-fitting PSDs. The black curve shows the observed PSD estimated via the FFT method. The purple curve corresponds to the best-fitting model, which consists of a power law (red noise; the pink dashed line) and a constant (measurement noise; the orange dotted line). The shaded regions correspond to the $1\sigma$ confidence intervals. Note that the confidence interval for the measurement noise is too small to be visible.}
    \label{fig:psd}
\end{figure}

An example of the GP modeling for the \textit{obs-ID} $0800840201$ is shown in Figure~\ref{fig:fit}. For this observation, the best-fitting ``mean'' function is a skewed Gaussian function (constant) for the QPO+DRW (noise) model. For the QPO+DRW, a QPO signal is prominent in the best-fitting PSD; for the noise model, the best-fitting PSD follows a power-law shape. The corresponding $\ln \mathcal{BF}=4.5$ (i.e., strong support for the QPO+DRW model), and the best-fitting period is $P_{\mathrm{QPO}}(1.78^{+0.17}_{-0.14})\times 10^4\ \mathrm{s}$. For the remaining two \textit{XMM-Newton} light curves, the corresponding $\ln \mathcal{BF}$ values are close to or smaller than zero. Hence, we focus on the \textit{obs-ID} $0800840201$. 

\begin{figure*}
    \centering
    \includegraphics[width=17cm]{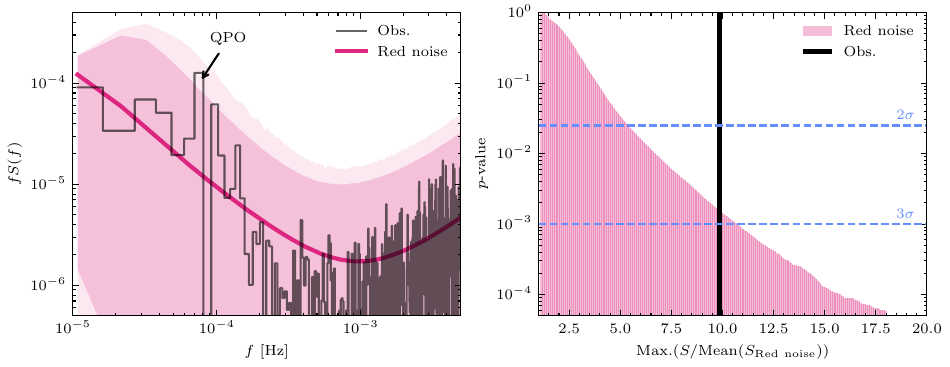}
    \caption{The FFT analysis of NGC~1566. Left: the FFT PSD of the observed light curve with obs-ID $0800840201$ (black curve). The pink curve is the average FFT PSD of the $5\times 10^5$ mock light curves (see text), with the shaded pink regions indicating its $3\sigma$ and $4\sigma$ confidence intervals. The QPO signal has a local $p$-value of $1.8\times 10^{-4}$ ($3.6\sigma$) for the period $P_{\mathrm{QPO}}=1.32\times 10^4\ \mathrm{s}$. Note that the y-axis is the product of the PSD and frequency. Right: the pink histogram represents the global $p$-value (i.e., the survival function, accounting for the ``look elsewhere'' effect) derived from our red-noise Monte Carlo simulations. The blue dashed lines indicate the $2\sigma$ and $3\sigma$ significance levels. The black line represents the observed QPO signal strength, resulting in a global $p$-value of $1.5\times 10^{-3}$ (i.e., $2.96\sigma$).} 
    \label{fig:fft}
\end{figure*}

To further justify our QPO detection, we performed the standard FFT analysis for the \textit{obs-ID} $0800840201$. We can use the standard FFT analysis because the light curve is evenly sampled. The PSD from the FFT analysis is shown in Figure~\ref{fig:fft}. We searched for QPO signals at frequencies $f<5\times 10^{-4}$ Hz, corresponding to periods $>2000$ s. We choose this limit because the observed variability is dominated by measurement uncertainties at $f>\sim 5\times 10^{-4}$ Hz (see Figure~\ref{fig:psd}). A QPO signal with the period of $P_{\mathrm{QPO}}=1.32\times 10^4\ \mathrm{s}$ is detected, and the period is roughly consistent with our GP fitting result. 

We use Monte Carlo simulations to assess the possibility ($p$-value) that the red noise could reproduce the observed (or even stronger) QPO signal. As a first step, we fit the FFT-based PSD with the following model, $S(f)=A(f/(10^{-4}\ \mathrm{Hz}))^{-\beta}+C$, where $A$ and $\beta$ are parameters characterizing the PSD of a red noise component, and $C$ denotes the contribution due to measurement errors. The best-fitting values for these parameters are obtained by maximizing the Whittle's likelihood \citep{Whittle1953}. The best-fitting parameters and their $1\sigma$ uncertainties are $A=1.40\pm 0.05$, $\beta=3.07\pm0.27$, and $C=(8.29\pm0.39)\times 10^{-4}$, respectively. The uncertainties are estimated by resampling the observed PSD via bootstrapping. Note that the inferred $\beta$ is steeper than the DRW PSD. The best-fitting PSD and its $1\sigma$ confidence interval are presented in Figure~\ref{fig:psd}. Given that the confidence interval is rather small, we neglect it for further analysis. We then use the best-fitting red-noise PSD to generate $5\times 10^5$ realizations of parent mock light curves following the methodology presented in \cite{TK1995}. To avoid red-noise leakage effects, the parent mock light curves are $40$ times longer and $5$ times denser than the observed one. For each parent mock light curve, we use the linear interpolation to obtain a mock light curve whose cadence and duration are identical to the observed one. We also add white noise to match the signal-to-noise ratio of the observed \textit{XMM-Newton} light curve. Then, we calculate the FFT-based PSD for each mock light curve following the same procedures applied to the observed one. The average of these $5\times 10^5$ mock PSDs, shown as the pink curve in Figure~\ref{fig:fft}, provides a good fit to the overall shape of the observed PSD. The shaded regions correspond to the $3\sigma$ and $4\sigma$ confidence intervals of our $5\times 10^5$ mock PSDs. Then, we can conclude that the local $p$-value---defined as the probability of the noise model reproducing the observed (or even stronger) QPO signal specifically around $P_{\mathrm{QPO}}=1.32\times 10^4$ s---is less than $0.018\%$. 

We must also take into account the ``look-elsewhere'' effect and calculate the corresponding global $p$-value as we had no strong prior constraint on the QPO period being around $P_{\mathrm{QPO}}=1.32\times 10^4$ s. As mentioned in Section~\ref{sec:results}, we restricted our search to frequencies $f<5\times 10^{-4}$ Hz, or periods $>2000$ s. This choice is motivated by two reasons. First, the expected high-frequency QPO for NGC~1566, based upon its SMBH mass, is $2000\sim 4000$ s. Second, at frequencies higher than $5\times 10^{-4}$ Hz, the PSD is dominated by measurement noise. To identify potential QPO signals, we calculate the ratio of the observed PSD to the mean mock PSD. The QPO frequency is defined as the frequency that maximizes this ratio (hereafter, $\rho_{\mathrm{max,obs}}$). For each mock PSD, we likewise calculate its maximum ratio to the mean mock PSD (hereafter, $\rho_{\mathrm{max,mock}}$). The survival function of these $5\times 10^5$ $\rho_{\mathrm{max,mock}}$ is shown in the right panel of Figure~\ref{fig:fft}. The vertical black line indicates $\rho_{\mathrm{max,obs}}$. The blue horizontal lines indicate the $p$-values for $2\sigma$ and $3\sigma$ confidence levels, respectively. The global $p$-value---the probability of the noise model reproducing the observed (or even stronger) QPO signal within the searched range of $10^5\ \mathrm{s}>P_{\mathrm{QPO}}>2000\ \mathrm{s}$---is simply the fraction of simulations where $\rho_{\mathrm{max,mock}}\geq \rho_{\mathrm{max,obs}}$, i.e., $0.15\%$. This corresponds to a significance level of $2.96\sigma$. We therefore identify an X-ray QPO candidate in the \textit{XMM-Newton obs-ID} $0800840201$ when NGC 1566 is in its $2018$ outburst. 

\section{QPOs in NGC~1566 and other SMBH accreting systems} \label{sec:results}
The QPO+DRW model can well fit the merged and rebinned \textit{XMM-Newton} light curve of NGC~1566 (Figure~\ref{fig:fit}). The best-fitting power spectral density (PSD) has a prominent peak beyond the red noise, corresponding to a QPO period of $P_{\mathrm{QPO}}=(1.27^{+0.07}_{-0.06})\times 10^4$ s ($1\sigma$). The combination of the two \textit{XMM-Newton} observations also allows the red noise at low frequencies ($\lesssim 10^{-5}$ Hz) to be well constrained. The Bayesian evidence for this model is $\ln(\mathcal{Z}_\mathrm{QPO+DRW})=-2181.6$. For comparison, we also fit the same \textit{XMM-Newton} light curve with the aforementioned noise model (i.e., by artificially fixing $P_{\mathrm{QPO}}$ to an extremely large value in Eq.~\ref{eq:full}), yielding $\ln(\mathcal{Z}_\mathrm{noise})=-2196.5$. Nearly identical Bayesian evidence is obtained when fitting with a standard DRW model. The natural logarithm of the Bayes factor between the QPO+DRW and noise models is therefore $14.9$, indicating strong support for the presence of a QPO. 

The measured QPO period for NGC~1566 can be compared with the high-frequency QPO scaling relation of BHXRBs. Using the SMBH mass compiled by \cite{Ochmann2024}, the SMBH mass from the $M_{\mathrm{BH}}$-sigma relation is $4\sim 8 \times 10^6\ M_{\odot}$. The high-frequency QPO scaling relation of BHXRBs \citep[i.e., $P_{\mathrm{QPO}}=(M_{\mathrm{BH}}/M_{\odot})/1862\ \mathrm{s}$; see, e.g.,][]{Remillard2006, Zhou2015} predicts a high-frequency QPO period of $2148.2\sim 4296.4$ s. The measured QPO period is a factor of $3\sim 6$ larger than this expected range. However, given the rather large uncertainties in $M_{\mathrm{BH}}$, we cannot rule out the possibility that the detected QPO candidate is an analog of high-frequency QPOs in BHXRBs. 

We combine our result with previously reported X-ray AGN QPOs to construct a new QPO sample (Table~\ref{tab:qpo}). This sample is smaller than those compiled by \cite{Smith2021} or \cite{Zhang2025} because we exclude sources that lack independent $M_{\mathrm{BH}}$ measurements (i.e., those derived from methods other than the high-frequency QPO scaling relation) or whose QPOs were not in X-ray. We also include high-frequency QPOs in BHXRBs and other black-hole systems. These sources can be classified as BHXRBs, TDEs, QPEs \citep[quasi-periodic eruptions; e.g.,][]{Miniutti2019, Giustini2020}, CL-AGN, and NLS1s (see Table~\ref{tab:qpo}). Several sources with QPEs are also classified as TDEs because there is a strong connection between QPEs and TDEs \citep[e.g.,][]{Nicholl2024}. For instance, \cite{Jiang2025} propose that QPEs are produced through collisions between a stellar-mass object (possibly formed in a prior AGN disk) and the newly formed misaligned TDE disk. BHXRBs often show different types of QPOs \citep[for a review, see e.g.,][]{Remillard2006}. As pointed out by \cite{Smith2021} and \cite{Zhang2025}, these QPOs in SMBH accretion systems\footnote{With the possible exceptions of 2XMM 123103.2+110648 \citep[which is claimed to be a low-frequency QPO analog; see][]{Lin2013} and GSN 069.} are analogs of high-frequency QPOs in BHXRBs \citep[e.g.,][]{Zhou2015}.

\section{Discussion} \label{sec:dis}
Our results in Section~\ref{sec:results} suggest that QPOs in SMBH accretion systems are frequently detected in NLS1s or TDEs. Our following discussion is based on the assumption that high-frequency QPOs are indeed preferretially detected in these types of systems. As pointed out by \cite{Smith2021}, this conclusion may suffer from selection biases. Specifically, most of the archival \textit{XMM-Newton} observations were not optimized for QPO searches, and the observing conditions (e.g., exposure time and cadence) vary significantly among sources, which can significantly affect the QPO detection efficiency \citep[e.g.,][]{Vaughan2005}. Nevertheless, as pointed out by \cite{Smith2021}, \cite{Vaughan2012} uniformly analyzed the X-ray variability of $104$ AGN with \textit{XMM-Newton} observations and found the robust QPO signal only in RE J1034+396 \citep[also see, e.g.,][]{Gierlinski2008, Jin2021}, which is an NLS1. Other QPO detections in AGN from new data have also predominantly been in NLS1s. While NLS1s tend to have smaller $M_{\mathrm{BH}}$ (hence shorter expected $P_{\mathrm{QPO}}$) than broad-line Seyfert 1s (BLS1s), some BLS1s with $M_{\mathrm{BH}}<10^7\ M_{\odot}$ lack QPO detections \citep[e.g., NGC 4593, NGC 7469, NGC 3227; see,][]{Vaughan2012}. In addition, QPOs seem not to be uncommon in TDEs \citep[e.g.,][]{Pasham2019, Masterson2025, ZhangWJ2025} with or without QPEs, with detections in over five sources among several dozen TDEs with high-quality X-ray observations \citep[e.g.,][]{Gezari2021}. Hence, it appears that QPOs are often found in NLS1s or TDEs. 

\begin{table*}
\caption{A summary of X-ray QPO properties}
\label{tab:qpo}
\centering
\begin{tabular}{lllcccc}
\hline \hline
Name & $P_{\mathrm{QPO}}\ \mathrm{[s]}$ & QPO reference & $M_{\mathrm{BH}}\ [M_{\odot}]$ & $\log(\lambda_{\mathrm{Edd}})$ & Mass reference & Type \\
\hline
NGC 1566 & $17800$ & this work & $10^{6.86}$ & $-1.2$ & (1) & CL-AGN \\
\hline
MCG-6-30-15$^{a}$ & $3670$ & (2) & $10^{6.60}$ & $-1.1$ & (1) & NLS1 \\
\hline
RE J1034+396 & $16900$ & (3) & $10^{6.74}$ & $\sim 1.0$ & (4) & NLS1 \\
\hline
MS 2254.9-3712$^{a}$ & $7200$ & (5) & $10^{6.6}$ & $-0.6$ & (6) & NLS1 \\
\hline
NGC 5506 & $16000$ & (7) & $10^{7.94}$ & $\sim -1.3$ & (8) & Obscured NLS1 \\
\hline
Mrk 766$^{a}$ & $6450$ & (9) & $10^{6.82}$ & $-1.25$ & (10) & NLS1 \\
\hline
1H 0707-495$^{a}$ & $3800$ & (11) & $10^{6.72}$ & $-0.3$ & (11) & NLS1 \\
\hline
NGC 1365$^{b}$ & $4566$ & (12) & $10^{6.85}$ & $-1.5$ & (13) & CL-AGN \\
\hline
1ES 1927+654 & $559$ & (14) & $10^{6.14}$ & $-0.4$ & (15) & CL-AGN/TDE \\ 
\hline
RX J1301.9+2747 & $1500$ & (16) & $10^{5.9}$ & $-1.0$ & (16) & QPE \\
\hline
GSN 069 & $32000$ & (17) & $10^6$ & $-0.8$ & (17) & QPE/TDE \\
\hline
2XMM J123103.2+110648 & $13680$ & (18) & $10^{5}$ & $-1.0$ & (19) & TDE \\
\hline
Swift J164449.3+573451 & $200$ & (20) & $10^{5.5}$ & $1.2$ & (21) & TDE \\
\hline
ASASSN-14li & $131$ & (22) & $10^{6.23}$ & $-0.6$ & (23) & TDE \\
\hline
H1743-322 & $0.006$ & (24) & $10^{1.08}$ & SPL & (25) & XRB \\
\hline
GRS 1915+105 & $0.009$ & (26) & $10^{1.09}$ & SPL & (27) & XRB \\
\hline
GRO J1655-40 & $0.003$ & (28) & $10^{0.8}$ & SPL & (29) & XRB \\
\hline
XTE J1859+226 & $0.005$ & (30) & $10^{0.89}$ & SPL & (31) & XRB \\
\hline
XTE J1650-500 & $0.004$ & (32) & $<10^{0.86}$ & SPL & (33) & XRB \\
\hline
XTE J1550-564 & $0.005$ & (28) & $10^{0.96}$ & SPL & (34) & XRB \\
\hline
\end{tabular}
\tablebib{a. the QPO was not detected according to \cite{Zhang2023}. b. the QPO was not detected by \cite{Gurpide2025}. c. the reported QPOs in RX J0437.4-4711, Ton S180, and Sgr A$^{\star}$ are excluded due to the lack of X-ray detections, and the QPOs in AT2020ocn \citep{Pasham2024} and AT2020afhd \citep{Wang2025} are excluded due to their extremely low-frequency and likely distinct physical origins. d. the reported QPOs in XMMU J134736.6+173403 \citep[Sy2;][]{Carpano2018}, 3XMM J215022.4-055108 \citep[TDE;][]{ZhangWJ2025}, NGC 5408 X-1 \citep[ultraluminous X-ray sources; ULX;][]{Strohmayer2007}, M82 X-1 \citep[ULX;][]{Pasham2014}, and 4U 1630-47 \citep[XRB;][]{Remillard2006} are not included due to the lack of independent $M_{\mathrm{BH}}$ measurements. e. for XRBs, the QPOs are high-frequency ones. f. the Eddington ratios for SMBH accreting systems are either obtained from their corresponding QPO references or estimated by this work. g. for XRBs, high-frequency QPOs are found in the SPL state with high Eddington ratios. (1)~\citet{Koss2022}; (2)~\citet{Gupta2018}; (3)~\citet{Xia2025}; (4)~\citet{Czerny2016}; (5)~\citet{Alston2015}; (6)~\citet{Grupe2010}; (7)~\citet{Zhang2025}; (8)~\citet{Papadakis2004}; (9)~\citet{Zhang2017}; (10)~\citet{Bentz2015}; (11)~\citet{Pan2016}; (12)~\citet{Yan2024}; (13)~\citet{Fazeli2019}; (14)~\citet{Masterson2025}; (15)~\citet{Li2022}; (16)~\citet{Song2020}; (17)~\citet{Miniutti2023}; (18)~\citet{Lin2013}; (19)~\citet{Ho2012}; (20)~\citet{Reis2012}; (21)~\citet{Miller2011}; (22)~\citet{Pasham2019}; (23)~\citet{Wevers2017}; (24)~\citet{Homan2005}; (25)~\citet{Nathan2024}; (26)~\citet{Remillard2004}; (27)~\citet{Reid2014}; (28)~\citet{Remillard2002}; (29)~\citet{Greene2001}; (30)~\citet{Cui2000}; (31)~\citet{Yanes-Rizo2022}; (32)~\citet{Homan2003}; (33)~\citet{Orosz2004}; (34)~\citet{Orosz2011}.}
\end{table*}

\subsection{Why are QPOs often found in TDEs or NLS1s? \label{subsec:origin}}
Both NLS1s and TDEs are characterized by high (peak) Eddington ratios \citep[e.g.,][]{Williams2018, Gezari2021}, presumably resembling the SPL state of BHXRBs---the only spectral state in which high-frequency QPOs in BHXRBs are detected \citep[e.g.,][]{Remillard2006, Belloni2012}. Theoretically, a high Eddington ratio alone may not be a sufficient condition to drive high-frequency QPOs. As noted in Section~\ref{sec:intro}, MHD simulations \citep[e.g.,][]{Dewberry2020} reveal that MHD turbulence can damp the $g$-mode oscillations, a key process of the promising diskoseismology model for high-frequency QPOs. It has been demonstrated that misaligned and/or eccentric accretion flows can excite \citep[e.g.,][]{Kato2004, Ferreira2008, Henisey2009} and sustain the $g$-mode oscillations, thereby producing high-frequency QPOs \citep[see, e.g.,][]{Dewberry2020, Musoke2023, Bollimpalli2024}. \cite{Smith2021} mention that the fundamental $g$-mode oscillations cannot well account for the observed QPO frequencies, although this conclusion relies on the rather large uncertainties in SMBH mass estimations. In addition, misaligned accretion may also excite higher-order $g$-mode oscillations with larger frequencies than the fundamental one or $p$-mode waves that are responsible for the observed QPOs \citep[e.g.,][]{Henisey2009}. We now interpret our results with the assumption that the diskoseismology model is valid and that misaligned and/or eccentric accretion is required to drive discoseismic oscillations. 

Misaligned and/or eccentric accretion is, in fact, a plausible scenario in TDEs because the newly formed accretion disk is often expected to be misaligned and/or eccentric \citep[e.g.,][]{Rees1988}, as the orbit of the disrupted star is rarely initially aligned with the spin of the central SMBH. The alignment timescale between the SMBH and the misaligned and/or eccentric disk in a TDE can range from tens of days to several years, depending upon the black-hole spin, accretion rate, black-hole mass, and the viscosity parameter $\alpha$ \citep[e.g.,][]{Franchini2016, Zanazzi2019}. For ASASSN-14li, the QPO signal is detectable for at least $\sim 450$ days, which should be treated as a lower limit of the alignment timescale; therefore, the SMBH spin cannot be near maximal \citep[see figure~12 of][]{Zanazzi2019} and the viscosity parameter $\alpha$ cannot be around unity \citep[see figure~10 of][]{Franchini2016}. In one \textit{Suzaku} and a series of \textit{XMM-Newon} observations of Swift J164449.3+573451, a QPO signal is detected in the \textit{Suzaku} and the first \textit{XMM-Newon} observation; if the non-detection of the QPO in subsequent \textit{XMM-Newon} data is not caused by the contamination from red noise or measurement error, the alignment timescale would be around $20$ days. Given that the black-hole mass of this source is likely much smaller than that of ASASSN-14li, the $20$-day alignment timescale is possible. As for 2XMM J123103.2+110648, the \textit{XMM-Newton} observations are too sparse to constrain the alignment timescale.  

In AGN, misaligned accretion disks can also form via mechanisms such as tilted gas inflows \citep[e.g.,][]{Rees1978, Volonteri2005}, and randomly oriented accretion events on sub-pc scales can readily power Seyfert galaxies \citep[e.g.,][]{King2007-fuel}. This leads to the question: why might NLS1s be particularly prone to having misaligned and/or eccentric accretion disks? We propose two possible explanations. First, misaligned accretion disks may be a distinctive feature of NLS1s, setting them apart from BLS1s. Theoretically, the misaligned accretion disk will eventually be aligned with the SMBH spin due to the Bardeen-Petterson effect \citep{Bardeen1975}. The timescale for this alignment \citep{Scheuer1996} is $\sim 10^5$--$10^6$ years for near-Eddington accretion \citep[e.g.,][]{Natarajan1998, Volonteri2005}. This youth scenario is supported by substantial evidence, including their characteristic properties such as low $M_{\mathrm{BH}}$, high $\lambda_{\mathrm{Edd}}$, and distinct large-scale environments \citep{Jarvela2017}, which suggest that the current episode of black-hole accretion was recently triggered \citep[e.g.,][]{Mathur2000}. Therefore, NLS1s with QPOs could be young AGN in which the accretion disks and SMBHs have not yet achieved alignment. Second, misaligned accretion disks may be common in AGN, but only those with high Eddington ratios (like NLS1s) can allow misalignment or eccentricity effects to propagate into the innermost regions. In summary, under the diskoseismology model, a misaligned accretion flow capable of inducing inertial oscillations in the innermost regions and driving high-frequency QPOs is plausible.

\subsection{The fuelling mechanism of CL-AGN\label{subsec:fueling}}
The detection of a QPO in NGC~1566 during its $2018$ outburst has interesting implications for our understanding of CL-AGN. As noted in Section~\ref{sec:intro}, NGC~1566 reached an Eddington ratio of $6\%\sim 12\%$ in its $2018$ outburst, and its optical spectra in the outburst show significant Fe~\textsc{II} emission \citep{Oknyansky2019}. It is therefore reasonable to speculate that, around the peak of the outburst, NGC~1566 reached a spectral state analogous to the SPL state in BHXRBs. Very recently, a QPO was also detected in another CL-AGN, 1ES 1927+654 \citep{Masterson2025}, with a large peak Eddington ratio of $\lambda_{\mathrm{Edd}}\simeq 1$ \citep[e.g.,][]{LiRC2024}. This source has been interpreted as a TDE \citep[e.g.,][]{Ricci2020} or a tilted gas inflow event, both of which imply a misalignment between the newly supplied gas and the SMBH spin. A QPO has been tentatively reported in the CL-AGN NGC~1365 \citep[which shows absorption and accretion-rate variations;][]{Jana2025} by \cite{Yan2024} but questioned by \cite{Gurpide2025}. Nevertheless, if the excitation mechanism of the QPO candidate in NGC~1556 is misaligned or eccentric accretion, our result supports the picture that the gas fuelling during its turn-on phase may be randomly orientated, potentially resulting in disks that are misaligned with the SMBHs and/or have a large eccentricity. Such a disk may precess and drive the possible ``near-periodic'' long-term X-ray fluctuations reported by \cite{Jana2021}. The misalignment or eccentricity effects can propagate inwards and induce inertial oscillations in the innermost regions, thereby producing high-frequency QPOs observed in the light curve of \textit{ovs-IDs} $0800840201$. For the other two \textit{XMM-Newton} observations, the X-ray luminosity drops by a factor of $\sim 4$ \citep[table 7 in][]{Jana2021} and the misalignment or eccentricity effects might be damped in low-$\lambda_{\mathrm{Edd}}$ cases. If QPOs are indeed more common in CL-AGN than other AGN, many CL events may be powered by misaligned or eccentric gas fuelling. 

Substantial evidence suggests that QPEs are produced through collisions between a stellar-mass object (possibly formed in a prior AGN disk) and the newly formed, misaligned/eccentric TDE disk \citep{Xian2021, Franchini2023, Tagawa2023, Linial2023, Jiang2025}. In this scenario, QPEs serve as a reliable indicator of misaligned/eccentric gas fuelling \citep[see also][who also allude to a similar idea]{Lyu2025}. Interestingly, \cite{Hernandez2025} reported QPEs in a newly ``turn-on'' AGN, $\mathrm{SDSS\ J133519.91+072807.4}$ \citep[a.k.a., ``Ansky''; but see][for the evidence of the alternative TDE possibility]{Zhu2025}, suggesting the presence of misaligned accretion in the ``turn-on'' phase of this AGN. The same mechanism may also be responsible for powering CL-AGN and creating a tearing disk to drive CL-AGN variability \citep[e.g.,][]{Raj2021}. We therefore speculate that, in CL-AGN, a low angular momentum gas inflow supplies a misaligned and/or eccentric accretion disk. The interaction between the newly formed misaligned and/or eccentric accretion disk and the pre-existing AGN disk can induce shocks, efficiently redistribute gas angular momentum, and greatly reduce the accretion timescale, thereby accounting for the short transition timescales in CL-AGN. A critical question remains: where is the gas that supplies the newly formed misaligned and/or eccentric accretion disks in CL-AGN? 

We propose that broad line regions (BLRs) may act as the gas reservoir, supplying misaligned and eccentric fuel in CL-AGN. This offers an alternative to models involking chaotic accretion from much larger scales \citep{Liu2021, Wang2024, Veronese2024}, which must overcome the angular momentum barrier to effectively fuel the central SMBH. \cite{Krolik1988} proposed $\sim 1\ M_{\odot}\mathrm{yr^{-1}}$ inflows from dusty tori due to the cloud-cloud collisions; the dusty inflowing clouds may be responsible for BLRs \citep{Wang2017}. We speculate that a similar mechanism could operate within the BLR itself. Dynamical modeling of some AGN, including the CL-AGN NGC 5548 \citep{Pancoast2014}, NGC 4151 \citep{Ramachandran2025}, and NGC~1566 \citep{Ochmann2024}, suggests that a fraction of BLR clouds can have eccentric orbits. While the eccentricity of most clouds is modest, it is quite possible that several clouds are highly eccentric and have very low specific angular momentum. Furthermore, the proposed existence of spiral arms in BLRs \citep[e.g.,][]{Du2023} could channel misaligned gas inflows towards the SMBH, analogous to gaseous inflows in galactic disks \citep[e.g.,][]{Kim2017}. BLR clouds with similar low angular momentum may also manifest as ``quasar rain'' and produce X-ray eclipsing events \citep{Elvis2017}, which are observed in CL-AGN like NGC~3516, NGC~4151, and NGC~6814 \citep[e.g.,][]{Kang2023, Lian2025}. Recent numerical simulations find that low angular momentum gas accretion can have resonance around the horizon of the SMBH and generate QPOs \citep{Dihingia2025}. Hence, CL-AGN may be powered by the combination of a stable conventional, accretion disk and variable, misaligned gas fuelling directly from BLRs. 

Can BLR clouds provide sufficient accreting masses? Given that the critical $\lambda_{\mathrm{Edd}}$ for CL-AGN is about $0.01$ \citep[e.g.,][]{Ricci2023}, the accreted mass during a typical CL-AGN flare can be estimated as $M_{\mathrm{acc}}\sim 0.01\dot{M}_{\mathrm{Edd}}t_{\mathrm{dur}}\sim 0.02\ M_{\odot}(M_{\mathrm{BH}}/[10^8\ M_{\odot}])(t_{\mathrm{dur}}/[1\ \mathrm{yrs}])$, where $\dot{M}_{\mathrm{Edd}}=1.3\times 10^{18}(M_{\mathrm{BH}}/M_{\odot})\ \mathrm{g\ s^{-1}}$ is the Eddington accretion rate and $t_{\mathrm{dur}}$ is the CL-AGN timescale. Note that the peak $\lambda_{\mathrm{Edd}}$ in CL-AGN outbursts (e.g., the $2018$ outburst of NGC 1566) can be much higher than $1\%$. The mass of a typical BLR is $\sim 10^4\ M_{\odot}$ \citep[e.g.,][]{Baldwin2003}, which is six orders of magnitude larger than $M_{\mathrm{acc}}$. The BLR is widely speculated to contain millions of BLR clouds \citep[e.g.,][]{Arav1997}. The typical velocity of a BLR cloud is close to the Kepler velocity $v_{\mathrm{c}}\sim \sqrt{GM_{\mathrm{BH}}/R_{\mathrm{BLR}}}\sim 6.8\times 10^3\ \mathrm{km\ s^{-1}}(L_{5100}/[10^{43}\ \mathrm{erg\ s^{-1}}])^{-0.25}(M_{\mathrm{BH}}/[10^8\ M_{\odot}])^{0.5}$, where $R_{\mathrm{BLR}}=9.5\times 10^{-3}\ \mathrm{pc}(L_{5100}/[10^{43}\ \mathrm{erg\ s^{-1}}])^{0.5}$ is the BLR location \citep[e.g.,][]{Bentz2009} and $L_{5100}$ is the monochromatic luminosity at rest-frame $5100\ \mathrm{\AA}$. The BLR sound speed is only $c_{\mathrm{s}}\sim 10\ \mathrm{km\ s^{-1}}$ because the gas temperature is $T_{\mathrm{c}}\sim 10^4\ \mathrm{K}$ \citep[e.g.,][]{Elvis2017}. One can estimate the typical size of a BLR cloud as $R_{c}\leq c_{\mathrm{s}}(2\pi R_{\mathrm{BLR}}/v_{\mathrm{c}})\simeq 3\times 10^{14}\ \mathrm{cm} (L_{5100}/[10^{43}\ \mathrm{erg\ s^{-1}}])^{0.25}(\lambda_{\mathrm{Edd}}/0.01)^{0.5}$, where $\lambda_{\mathrm{Edd}}\simeq 10L_{5100}/L_{\mathrm{Edd}}$ with $L_{\mathrm{Edd}}$ represents the Eddington luminosity. Given the typical BLR gas number density of $n_{\mathrm{H}}\sim 10^{11}\ \mathrm{cm^{-3}}$ \citep[e.g.,][]{Elvis2017}, the mass of a BLR cloud is $M_{\mathrm{c}}=4\pi R_{\mathrm{c}}^3 n_{\mathrm{H}}m_{\mathrm{H}}/3\lesssim 0.009\ M_{\odot} (L_{5100}/[10^{43}\ \mathrm{erg\ s^{-1}}])^{0.75}(\lambda_{\mathrm{Edd}}/0.01)^{1.5}$. Therefore, the infall of just $N_{\mathrm{f}}\sim 1$---$10$ clouds on extremely eccentric and misaligned orbits---possibly triggered by cloud collision interactions \citep{Krolik1988}, spiral arms \citep{Du2023}, or condensation of disk winds \citep[][]{Elvis2017}---could supply sufficient accretion energy to power the ``turn-on'' phase of a CL-AGN. 

To produce UV/optical flares on the observed timescales, these clouds should reach tens of Schwarzschild radii and form a compact misaligned gas inflow/disk. The BLR cloud will be significantly reshaped by the tidal force. In fact, the tidal force at the BLR is $\sim GM_{\mathrm{BH}}M_{\mathrm{c}}R_{\mathrm{c}}/R_{\mathrm{BLR}}^3\leq (\pi R_{\mathrm{c}}^2)P_{\mathrm{c}}$, where $P_{\mathrm{c}}=n_{\mathrm{H}}m_{\mathrm{H}}c_{\mathrm{s}}^2$ is the gas pressure of the cloud. Unlike stars, the BLR clouds are not confined by self gravity as their masses are smaller than the Jeans mass. Instead, a long-standing hypothesis is that BLR clouds are confined by external pressure, such as the background magnetic field with the strength of $B(R_{\mathrm{BLR}})\simeq \sqrt{8\pi P_{\mathrm{c}}} \sim 1\ \mathrm{G}$ \citep{Rees1987}. At the BLR, the tidal force per unit area is close to the confinement pressure. As several clouds with low angular momentum fall into the SMBH, the tidal force will compress and stretch the clouds, even break them into a number of extended filaments. If the angular momentum is sufficiently small, the trajectory is extremely eccentric with the periapsis ($R_{\mathrm{p}}$) to be close to the SMBH. Then, these bounded and stretched filaments may collide, add gas to the pre-existing disk, redistribute gas angular momentum, and rapidly fuel the SMBH after several orbits. 

In our proposed scenario, the CL-AGN timescale is determined by the cloud inflow timescale. If the angular momentum is sufficiently small, the timescale is essentially the free-falling timescale from the BLR, i.e., $\tau_{\mathrm{ff}}=\sqrt{R_{\mathrm{BLR}}^3/(GM_{\mathrm{BH}})}\simeq 1.4\ \mathrm{yrs} (M_{\mathrm{BH}}/[10^8\ M_{\odot}])^{-0.5}(L_{5100}/[10^{43}\ \mathrm{erg\ s^{-1}}])^{0.75}=1.5\ \mathrm{yrs} (L_{5100}/[10^{43}\ \mathrm{erg\ s^{-1}}])^{0.25}(\lambda_{\mathrm{Edd}}/0.01)^{0.5}$. This timescale is insensitive to $L_{5100}$ or $M_{\mathrm{BH}}$ for fixed $\lambda_{\mathrm{Edd}}$ and is much shorter than the viscous timescale of an extended, conventional disk around the SMBH. Hence, it can reconcile with the observed changing-look timescale and its lack of dependence upon AGN properties \citep[e.g.,][]{Wang2024}. Moreover, our proposed gas fuelling mechanism should play a minor role in high Eddington-ratio AGN because the corresponding mass accretion rate of the conventional accretion disk would be large. This may explain why CL-AGN have $\lambda_{\mathrm{Edd}}\sim 0.01$ because our proposed gas fuelling can ignite SMBH activities with the accretion rate of $\sim M_{\mathrm{c}}/\tau_{\mathrm{ff}}\sim 0.02\dot{M}_{\mathrm{Edd}} (N_{\mathrm{f}}/10)(M_{\mathrm{BH}}/[10^8\ M_{\odot}])^{-1}(L_{5100}/[10^{43}\ \mathrm{erg\ s^{-1}}])^{1.5}$.

\section{Summary} \label{sec:sum}
In this work, we have searched for the QPO signal in the low-mass CL-AGN NGC~1566. Our results can be summarized as follows. 
\begin{itemize}
    \item We have detected a QPO candidate with the period $P_{\mathrm{QPO}}=(1.78^{+0.17}_{-0.14})\times 10^4\ \mathrm{s}$ in the $2018$ outburst of NGC 1566. This X-ray QPO likely represents an analog of the high-frequency QPOs in BHXRBs. 
    \item AGN X-ray QPOs are only found in NLS1s, QPEs, TDEs, and now in the CL-AGN NGC~1566. We have speculated that the high-detection rate of QPOs in NLS1s can be understood if NLS1s are young AGN and the accretion disk has not yet been aligned with the SMBH, similar to the newly formed accretion disks in TDEs. 
    \item The QPO detection in NGC~1566 suggests that the gas fuelling in CL-AGN may be misaligned and extremely eccentric BLR clouds. Such gas can form a compact accretion disk around the SMBH. The scenario can also easily explain the CL-AGN timescale, which is about two orders of magnitude smaller than the viscous timescale of an SSD. 
\end{itemize}

Future X-ray timing observations of NLS1s and CL-AGN can verify our scenario by testing the following prediction: QPOs (and potentially QPEs) should be detectable in CL-AGN and young NLS1s, provided these signals are not diluted by red noise. Alternatively, microlensing of CL-AGN can probe the accretion-disk sizes and test our proposed misaligned gas fuelling scenario, which involves a misaligned, compact accretion disk.

%% Please use the acknowledgment and contribution environments. This will 
%% be anonomyized when the "anonymous" style option is used. 
\begin{acknowledgements}
  We thank Du Pu and Min Du for the discussion of spiral arms in broad line regions and Wenjuan Liu for the inflows in AGN, Shanshan Weng for extracting the \textit{NuSTAR} data, and the referee for useful comments that improved this manuscript. M.Y.S. and S.Y.Z. acknowledge support from the National Natural Science Foundation of China (NSFC-12322303), the National Key R\&D Program of China (No.~2023YFA1607903), and the Fundamental Research Funds for the Central Universities (20720240152). J.F.W. acknowledges support from the National Natural Science Foundation of China (NSFC-12221003) and the National Key R\&D Program of China (No.~2023YFA1607904). Y.Q.X. acknowledges support from the National Natural Science Foundation of China (NSFC-12025303). This manuscript benefited from grammar checking by \texttt{DeepSeek} and \texttt{Grammarly}. We acknowledge the use of the \textit{XMM-Newton} Science Archive (XSA) and the \texttt{SAS} software. 
\end{acknowledgements}

\bibliographystyle{aa}
\bibliography{ref.bib}

\end{document}